\documentclass[letterpaper, 10 pt, conference]{ieeeconf}  %

\IEEEoverridecommandlockouts                              %

\usepackage{algorithm,algorithmic}
\usepackage{amsmath}
\usepackage{amssymb}  %
\usepackage{todonotes}
\usepackage{subfig}
\usepackage{mathtools}
\usepackage{comment}
\usepackage{siunitx}

\newcommand{\yref}{y_t^{\text{ref}}}
\newcommand{\yrefm}{y_t^{\text{ref},m}}

\title{\LARGE \bf
Robust nonlinear set-point control with reinforcement learning
}

\author{Ruoqi Zhang$^{1}$ , Per Mattsson$^{1}$  and Torbjörn Wigren$^{1}$ 
\thanks{*This research was financially supported by the project \emph{NewLEADS - New Directions in Learning Dynamical Systems} (contract number: 621-2016-06079), funded by the Swedish Research Council.}%
\thanks{$^{1}$Ruoqi Zhang, Per Mattsson and Torbjörn Wigren are with the Department of Information Technology, Division of Systems and Control,
        Uppsala University, SE-751 05 Uppsala, SWEDEN.
        }%
}
\def\layersep{2.5cm}

\begin{document}

\maketitle
\thispagestyle{empty}
\pagestyle{empty}

\begin{abstract}
There has recently been an increased interest in reinforcement learning for nonlinear control problems. However standard reinforcement learning algorithms can often struggle even on seemingly simple set-point control problems. This paper argues that three ideas can improve reinforcement learning methods even for highly nonlinear set-point control problems: 1) Make use of a prior feedback controller to aid amplitude exploration.  2) Use integrated errors. 3) Train on model ensembles. Together these ideas lead to more efficient training, and a trained set-point controller that is more robust to modelling errors and thus can be directly deployed to real-world nonlinear systems. The claim is supported by experiments with a real-world nonlinear cascaded tank process and a simulated strongly nonlinear pH-control system.

\end{abstract}

\section{Introduction}

Set-point control is a central standard functionality in industry and academia, with a vast number of studied solutions ranging from standard PID control \cite{astrom1970introduction, mcmillan2014tuning, ziegler1942optimum}, and model predictive control (MPC)  \cite{de2005constrained, karamanakos2020mpc} to tailored nonlinear adaptive methods \cite{krstic1995nonlinear}. To reduce the need for manual interaction and to improve performance further, reinforcement learning (RL) \cite{sutton2018reinforcement} is being introduced with the goal of learning optimal nonlinear control strategies, i.e. to perform optimal nonlinear adaptive control. The rationale behind this  is the success of RL as function approximators in a wide range of simulated tasks \cite{silver2016masteringgo, hessel2018rainbow, vinyals2019grandmaster}. However, there are still challenges when it comes to learning a feedback control policy for real-world systems \cite{dulac2019challenges}.
For example, the model-free RL methods, such as policy gradient methods \cite{sutton2018reinforcement, schulman2017proximal}, often need a large amount of training data and extensive exploration. The consequences can often be time-consuming training, and the exploration of the state space may lead to unsafe control actions. For this reason, a simulated environment is often used for training. However, there is no perfect model. The controller thus needs to be robustly trained with respect to varying system dynamics and unknown effects like occasional disturbances.

In this paper, a number of problems with standard RL algorithms are first isolated by the evaluation of a laboratory cascaded dynamic tank process \cite{wigren2006recursive-water-tank}. To ensure that the policy can handle varying set-points, the multi-goal framework from \cite{plappert2018multigoal} is utilized. It is stressed that conventional PID control could be used to solve this relatively simple problem, however, the main point here is that a reliable RL algorithm should also be able to handle seemingly simple problems like this. If the RL method cannot handle simple dynamic systems it is likely that the problems will amplify for highly nonlinear dynamic systems that are hard to model and where RL becomes one of few design alternatives. The main purpose of the paper is hence not to compare RL and conventional feedback control, it is rather to isolate and provide enhancements that allow standard RL algorithms to operate more efficiently and robust on highly nonlinear feedback control systems.

One observed drawback of standard RL methods is that they take a very long time to learn a policy, and may even fail on simple set-point control problems. As discussed in the paper, one problem is that because of the training on multiple set-points the importance of efficient exploration is increased.  If common  stochastic Gaussian policies are used for this purpose, the exploration of different amplitude domains is far too unlikely, and therefore the training process takes much longer time than necessary. 
The first contribution of the paper therefore applies an untuned proportional-controller (P-controller) as a prior controller to aid the exploration. 
After the modification, the policy learns how to track different set-points much more efficiently. But it is still time-wasting to collect samples on a real-world system. 
Thus, to facilitate deployment, the idea of training in simulation first and then using the policy in the real-world system is proposed. 
To handle the discrepancy between the real-world system and the trained simulated model, 
the standard approach of integrating control is applied by feeding the integrated error into the policy. 
However, even with the integrated error, simply training the policy on one fixed model is not enough since the optimal policy can instead find the optimal gains for this specific model. Thus, to force the RL policy to learn how to use the integrator a model ensemble \cite{rajeswaran2016-epopt, lim2013robust_mdp} is used for simulated training.
In summary, the paper argues that a combination of these three, relatively simple, adaptations of standard RL-methods are important for the training of a set-point controller and that removing one of these parts degrades either the efficiency of the training process or the robustness of the learned controller.

The proposed method is evaluated on the real-world cascaded double tank system and on a highly nonlinear pH-control problem in simulation. 
The pH cannot be well controlled by PID methods but the proposed RL method performs well. 
These examples show that the method can succeed in learning a robust policy that achieves time-varying set-point tracking, is insensitive to model errors and is also robust against unmodelled dynamics, even if the controlled dynamic system is so nonlinear that conventional feedback control becomes hard.

The idea of using a prior controller to aid RL is used in \cite{silver2018residual, johannink2019RRL},
where the prior controller is 
in the form of ad-hoc algorithms, proportional feedback, or model-based predictive control in a {\em fixed-goal} setting.  In \cite{zhang2021residualrl}, on the other hand, the focus is on exploration for  {\em multi-goal} set-point control, and it is shown that untuned PI-controllers are often sufficient to significantly improve the exploration of RL methods. 
In \cite{lawrence2020optimal_pid, carlucho2020adaptive_mimo_pid, lee2020reinforcement_adaptive_pid}, RL algorithms are used for tuning of the gains of PID controllers. The present work is fundamentally different in that no gains of the PI controller are tuned, the PI controller is fixed and rather used to aid and enhance exploration. The nonlinear control problem is solved directly by the aided RL algorithm. Classic tuning rules for PID controllers can be found in \cite{mcmillan2014tuning, ziegler1942optimum}, while an automatic tuning method deployed widely by Foxboro Inc. is available in \cite{aastrom1984automatic}. That method explores the system dynamics by triggering a limit cycle.

The paper is organized with a presentation of the proposed method in Sections~\ref{sec:problem_statement}-\ref{sec:summary}. Experimental results, motivating and validating the efficiency of the proposed method appear in Section~\ref{sec:exp}. Conclusions and further research directions close the paper in Section~\ref{sec:conclusion}.

\section{Problem Statement}
\label{sec:problem_statement}

Consider a discrete-time nonlinear state space model
\begin{align}
\label{eq:sp:model}
x_{t+1} &= f_p(x_t, u_t, w_t) \\
y_t &= h_p(x_t)
\end{align}
where $f_p$ and $h_p$ are unknown functions parameterized by $p$, $x_t \in \mathbb{R}^{n_x}$ is the current state of the system, $u_t \in \mathbb{R}^{n_u}$ is the input, $y_t \in \mathbb{R}^{n_y}$ is the output, and $w_t \in \mathbb{R}^{n_w}$ is the system noise. The task considered in the paper is to use RL-methods to train a state feedback set-point controller to track a piecewise constant set-point $\yref$.
To handle multiple set-points in RL, the multi-goal RL framework in \cite{plappert2018multigoal} is considered.
The aim is thus to find a controller
\begin{align}
u_t = g_{\theta}(x_t, \yref),
\end{align}
where $\theta$ is a parameter vector, that maximizes the discounted return

\begin{align}
\label{eq:return}
    G_t = \sum_{k=0}^{\infty} \gamma^k r_{t+k}
\end{align}
where $\gamma \in [0,1]$ is the discount factor and $r_t=R(x_t, u_t, \yref)$ is the reward function at time step $t$.  

As will be seen, state-of-the-art RL-methods like PPO struggle with this problem even in seemingly simple cases such as a cascade water tank system. To overcome the shortcomings of RL-methods, two simple extensions  are proposed -- making use of an integrator  to eliminate static errors and making use of a prior controller to improve the excitation of the system during the learning phase.

\section{Integrator and Robust Policy Optimization}
\label{sec:integrator:me}
With a simulation environment, training data can be collected cheaply and safely. However, since the simulation never describes the real-world system perfectly, there is a risk that RL will learn a controller that accounts also for errors in the simulation model \cite{rajeswaran2016-epopt}.

In set-point control, it is well-known that a pure state-feedback controller will result in a static error if the true gains of the system are not perfectly known. It is also well-known that this problem can be overcome by making use of integrating control \cite{aastrom2006advanced}. Using integrated errors is however {\it not} standard in the RL field. Here it will therefore be explored how this idea can be incorporated into the RL-framework.

Define the control error 
\[
\varepsilon_t = \yref - y_t.
\]Then the (discrete-time) integrated control error can be evaluated as
\begin{align}
    \label{def:integrator}
    z_t = z_{t-1} + \varepsilon_t .
\end{align}

In the literature on adaptive control, there are several results on tuning PID-controllers in a model-free manner \cite{mcmillan2014tuning, ziegler1942optimum, aastrom1984automatic}. However, these fix the structure of the policy to be linear in $\varepsilon_t$ and $z_t$. In this paper, the goal is rather to combine the power of integrating control with the generality of RL in the regulation of nonlinear dynamic systems. To achieve this, the extended state vector 
\begin{align}
    \tilde{x}_t = \begin{bmatrix}x_t^T & z_t \end{bmatrix}^T.
\end{align}
can be used to train a policy on the form $\pi_{\theta}(\tilde{x}_t, \yref)$ using standard RL-methods. 

\begin{figure}[htbp]
    \subfloat[]{
    \label{fig:no-me:model}
    \includegraphics[width=0.45\textwidth]{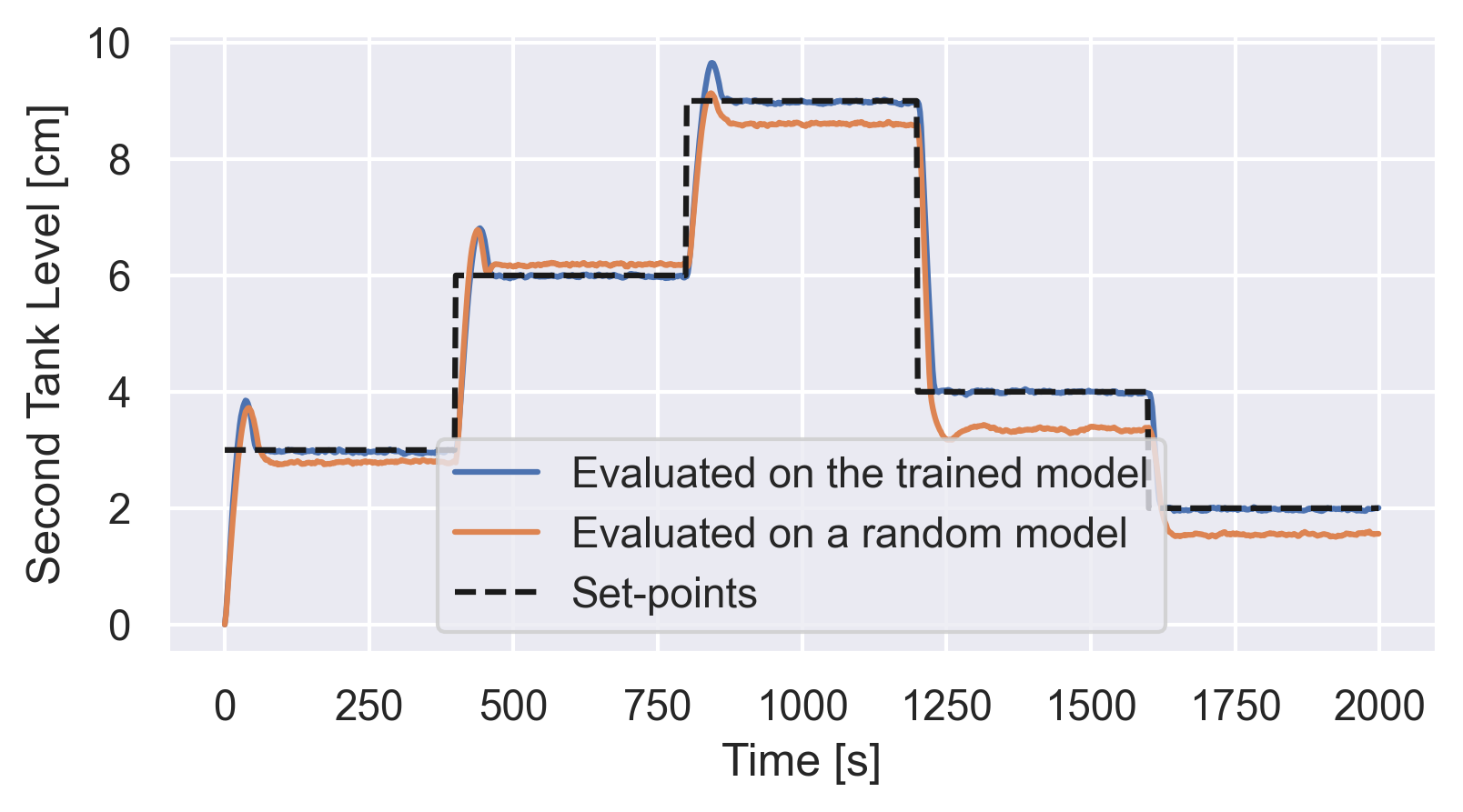}
    } \\
    \subfloat[]{
    \label{fig:no-me:policy}
    \includegraphics[width=0.46\textwidth]{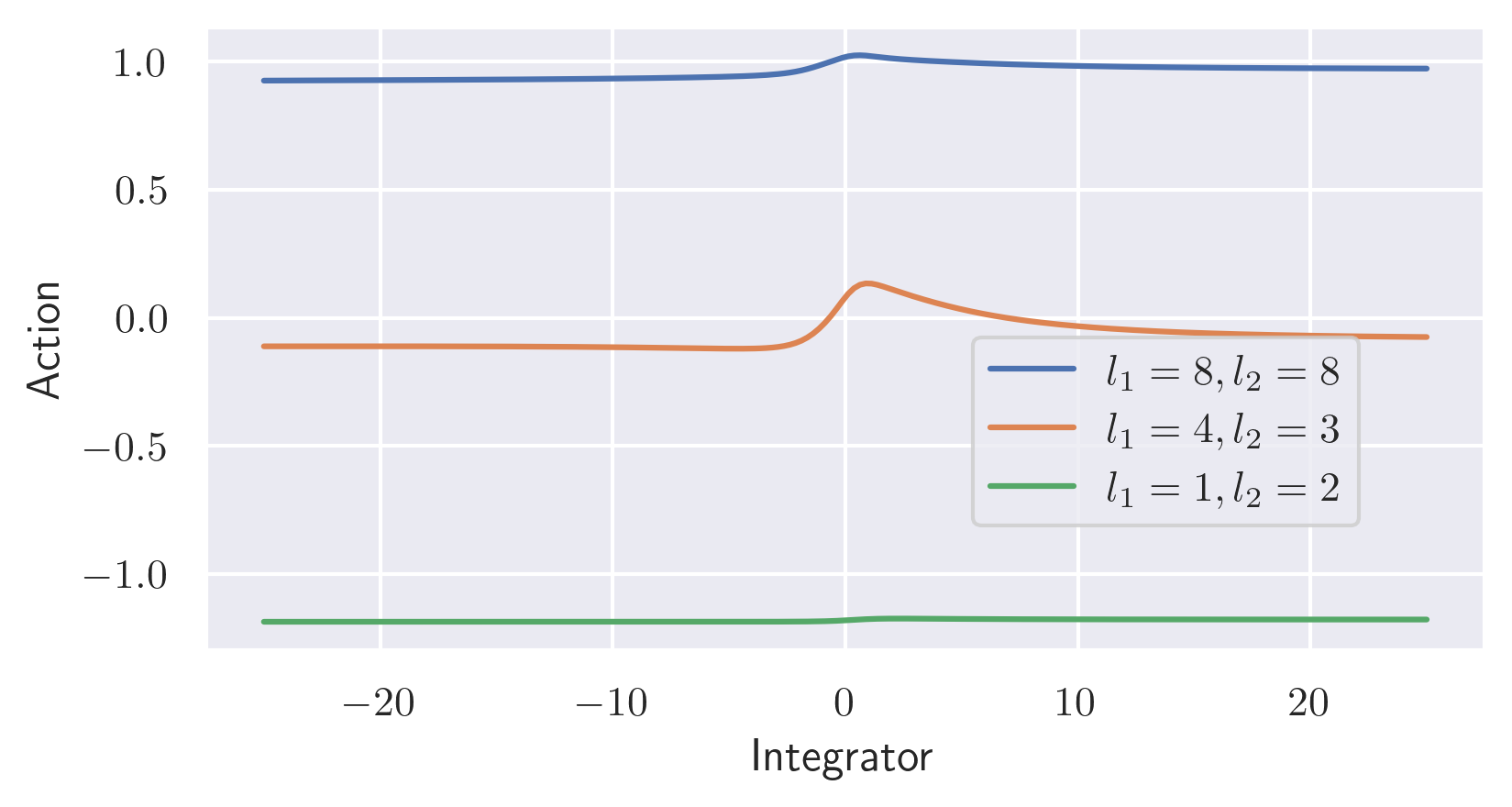}
    }
    \caption{Experiments on the RL-algorithm PPO with integrator but without model ensemble for the water tank level control task.(\emph{a}) Performance of controller evaluated on the trained model dynamics and a random model in simulation to track $5$ different set-points. (\emph{b}) The output of the trained controller for three situations with different value of water tank levels $(l_1, l_2)$ with different integrator values ($z_t$ in \eqref{def:integrator}).   }%
    \label{fig:res:no-me}%
\end{figure}

However, this is not enough. Suppose the controller is trained on a fixed model $f_p$. In that case, the RL-methods can and typically will find a controller that adjusts the gains depending on the set-points, instead of directly making use of the integrated error. That is, the resulting control signal will not depend on the integrated error. This situation is illustrated in Figure~\ref{fig:res:no-me}.

To force the policy to learn how to use the integrator and to handle the uncertainties about the system, it is proposed to use an ensemble of simulation models:
\begin{align}
    \mathcal{F}_\mathcal{P} = \{f_p \mid p \in \text{supp}(\mathcal{P})\},
\end{align} 
where $\mathcal{P}$ is some distribution of possible parameters, and $\text{supp}(\mathcal{P})$ is the support of this distribution. By training the policy on this model ensemble, instead of on a specific simulation model, the resulting policy will be more robust to model mismatch, see e.g. \cite{rajeswaran2016-epopt} and \cite{zhu2020transfer} and this will force the controller to utilize the integrator. 

Standard RL-methods, like \cite{schulman2017proximal, schulman2015trustregion}, can easily be extended to the ensemble case, see e.g. \cite{rajeswaran2016-epopt, kurutach2018-me-trpo}. Note that the controller does not know the parameters $p\in \mathcal{P}$, so it has to learn a control law that works well for all possible parameters.

\section{Aided Exploration in the Amplitude Domain}

\label{sec:efficient_exploration}
Exploration is generally a key to devising efficient RL algorithms. In the setting of model ensembles and multiple set-points, this is even more important. In control theory, this is known as excitation of the system dynamics \cite{astrom2008adaptive}.

In RL-methods, an often used way to achieve excitation is to add white probing noise to the controller. For example, in PPO the controller is defined via a Gaussian distribution $\pi_{\theta}(u_t \vert \tilde{x}_t, \yref)$ with mean $g_{\theta}(u_t \vert \tilde{x}_t, \yref)$ and some state-dependent variance.
While this is an efficient way of introducing excitation of different signal frequencies, it does by itself not give much exploration in the amplitude domain since the probing noise does not move the controlled variable in a systematic way \cite{wigren1990wiener}. In the set-point control setting, the exploration of amplitudes is however very important.

In the adaptive control literature, a common way to achieve exploration of amplitudes is to regularly change the set-point $\yref$ to force the controlled variable to move \cite{astrom2008adaptive}. 
When the controller is restricted to be e.g. a PI-controller changing the set-point will directly influence the actions and move the controlled variable to a different average level in a systematic way, thus exciting different amplitudes. However, when a general neural network $g_{\theta}$ is used, the training starts from a controller that chooses random actions independently of the current set-point. For this reason, it can take a long time before the RL-algorithm starts to explore relevant states.

In \cite{zhang2021residualrl}, it was noted that even with very limited knowledge about the environment, it is typically still easy to design a deterministic prior controller $\kappa(\tilde{x}_t, \yref)$ that at least will move the controlled variable in the general direction of the set-point. By starting the training process from such a prior policy, the RL policy will explore the environment in a much more efficient way. One way to achieve this is to consider a stochastic Gaussian policy on the form
\begin{equation} 
u_t \sim \pi_\theta(\kappa(\tilde{x}_t, \yref) + g_\theta(\tilde{x}_t, \yref) , \sigma_\theta^2 )  ,
\end{equation}
where $\theta$ is the parameter vector that should be trained and $\kappa$ is the prior policy. 
So how should this prior policy be chosen? Here the ideas from PI-control are again useful.
In many interesting environments, even an un-tuned PI-controller that results in stable operation will guide the controlled variable towards the set-points. 
\section{Summary of the Proposed Method}
\label{sec:summary}
The proposed method is thus a combination of three key ideas:
\begin{enumerate}
    \item Make use of an ensemble of simulation models to improve robustness to uncertainties in the parameters; see Section~\ref{sec:integrator:me}.
    \item Include the integrated control error in the state vector to find a policy that can track the set-point when the parameters change or that can handle unmodelled disturbances; see Section~\ref{sec:integrator:me}.
    \item Start from a prior controller to guide the policy to approach the desired set-point $\yref$ and thus assist the exploration, see Section~\ref{sec:efficient_exploration}.
\end{enumerate}
While each of these ideas is a simple extension of the standard multi-goal RL framework, the combination of them is important. Without the prior controller, algorithms like PPO may require a very long learning time to handle the multi-goal case, since the initial training data is not very informative. Without the integrator, there is no fixed policy that can handle changing parameters and unmodelled disturbances, a fact that would prevent safe real-world deployment.
Without robustness from the model ensemble, the trained policy can ``overfit'' the simulation environment, and the trained policy may not work well on the real-world target. For example, if the policy is trained in a fixed environment perfect tracking can be achieved without the integrated control error, and thus it may never learn to utilize this extra information. This is already shown by example in Section~\ref{sec:integrator:me}. 

The implementation of the method is straightforward. Given the user inputs,  the prior controller  $\kappa$, the desired set-points distribution $\mathcal{G}$, and the model ensemble $\mathcal{F}_\mathcal{P}$, the experience is collected by resetting the set-point and model at the beginning of each iteration. 
The policy is then directly trained by a model-free deep-RL method such as PPO. The summary of the method is shown in Algorithm~\ref{algo:PIME}.
\begin{algorithm}
 \caption{PIME for robust policy optimization}
 \begin{algorithmic}[1]
 \label{algo:PIME}
 \renewcommand{\algorithmicrequire}{\textbf{Input:}}
 \renewcommand{\algorithmicensure}{\textbf{Output:}}
 \REQUIRE Policy $\pi_\theta$ with a prior controller $\kappa$, the desired set-points distributions $\mathcal{G}$, model ensembles $\mathcal{F}_\mathcal{P}$
  \FOR {iteration=$1,2,\dots$,}
  \STATE Collect data with $M$ different set-points over $M$ different models
  \FOR{$m=1,2,\dots,M$}
    \STATE Sample $\yrefm$ from $\mathcal{G}$, sample $p$ from $\mathcal{P}$ uniformly and get the model $f_p$
    \STATE Reset the dynamic system $f_p$, randomly initialize the state $x_0$ and the integrator $z_0 = 0$
    \FOR{$t=0\dots T-1$,}
    \STATE Construct the extended state $\tilde{x}_t = \begin{bmatrix}x_t^T & z_t \end{bmatrix}^T$
    \STATE Interact with
    $f_p$  with the input $u_t \sim \pi_\theta(\kappa(\tilde{x}_t,y_t)+g_\theta(\tilde{x}_t, 
    \yrefm), \sigma_\theta^2)$
    \STATE Get next state $x_{t+1}$ from $f_p$ and the reward signal $r_t$ from reward function 
    \STATE Update the integrator $z_{t+1}$ and construct next embedded state $\tilde{x}_{t+1}=\begin{bmatrix}x_{t+1}^T & z_{t+1} \end{bmatrix}^T$
    \STATE Store $(\tilde{x}_t, \yrefm, u_t, \tilde{x}_{t+1}, r_t)$ and update $\tilde{x}_{t} = \tilde{x}_{t+1}$
    \ENDFOR 
  \ENDFOR
  \STATE Use an RL algorithm to optimize $\theta$, e.g., PPO
  \ENDFOR
 \end{algorithmic} 
\end{algorithm}

\section{Experiments}\label{sec:exp}
The proposed method was evaluated on two tasks. The first task is a laboratory cascaded water tank level control system. The policy was trained using simulations but evaluated also on a real-world cascaded water tank. The second task is to control the pH level in a strongly nonlinear wastewater treatment plant.
Again it is {\it stressed} that the purpose of the paper is not to advocate RL-based control for the cascaded tank process, the idea is rather to propose enhancements to make RL feedback control at least comparable to standard feedback control. The very hard  pH-control problem then shows how the enhancements enable efficient RL-based set-point control when PID control falls short.

As described in Section~\ref{sec:summary}, it is not possible to get perfect tracking over a range of parameters without the use of the integrated error together with the use of model ensembles. Thus, the methods that will be compared are:
\begin{itemize}
    \item model ensembles with PPO (IME-PPO): Uses the integrator and model ensembles.
    \item IME-PPO with a prior controller (PIME-PPO): 
    As prior controller an un-tuned PI-controller $\kappa(\tilde{x}_t, \yref) = -K_1\varepsilon_t + K_2 z_t$ is used.
\end{itemize}
Both methods are trained with five different random seeds, using the same hyperparameters for each task. The hyperparameters are tuned following the guidelines in \cite{andrychowicz2021matters} and the code is based on the library for deep reinforcement learning \cite{elegantrl}. 
All experiments are run on a MacBook Pro (16-inch, 2019) with 2,3 GHz 8-Core Intel Core i9. 
Detailed information about the hyperparameters and the neural network structure can be found in the Appendix.
The experiments are designed to answer the following questions:
\begin{enumerate}
    \item How well does the trained policy handle parameter changes?
    \item How does the policy trained in simulations work on the real-world system? 
    \item Is the policy robust to unmodelled effects?
\end{enumerate}

\begin{figure}[ht]
    \centering
    \subfloat[]{
    \label{fig:real-tank-pic}
    \includegraphics[width=0.2\textwidth]{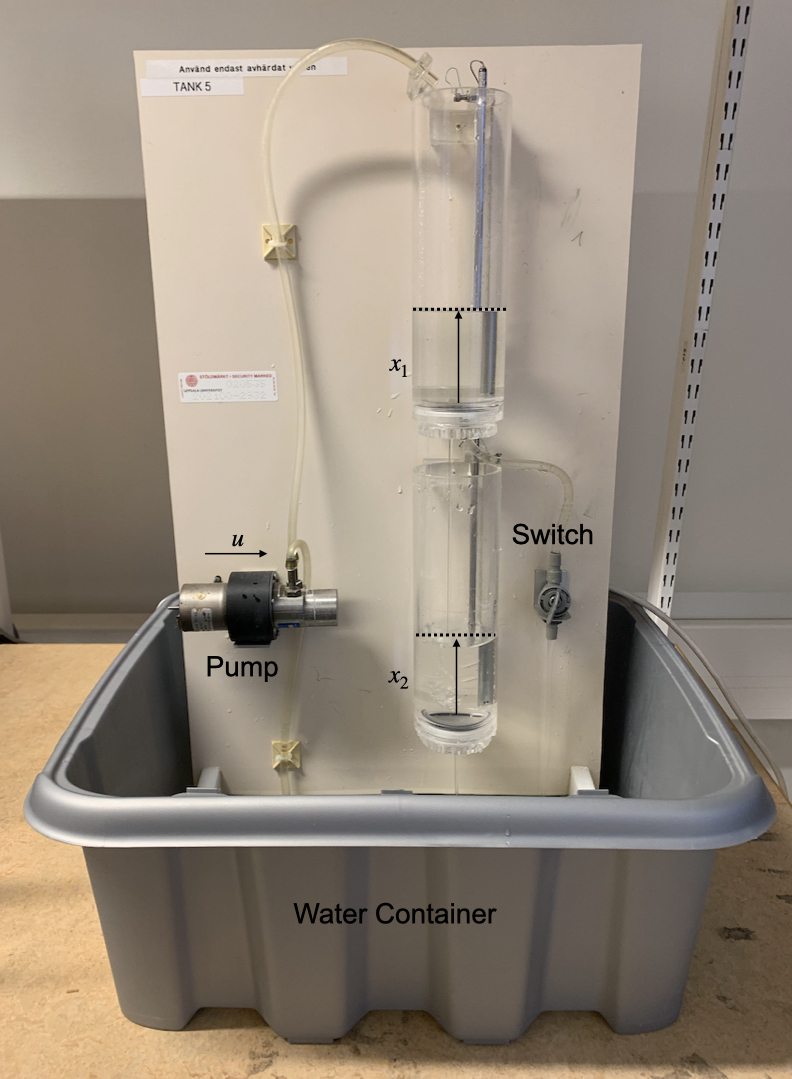}
    }
    \subfloat[]{
    \label{fig:ph-pic}
    \includegraphics[width=0.25\textwidth]{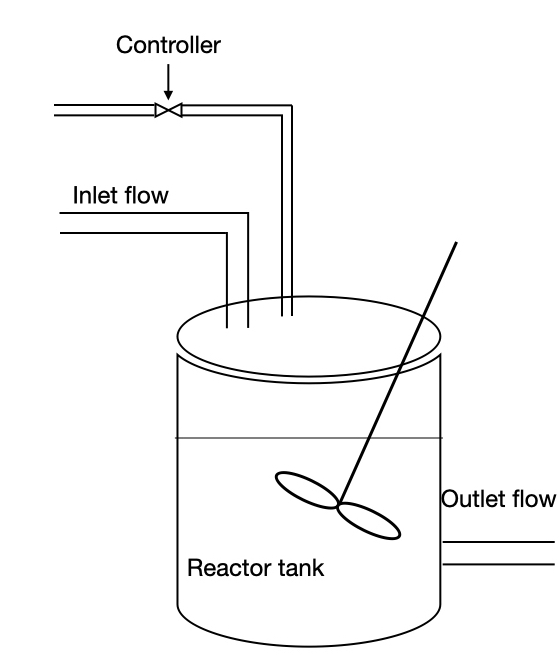}
    }
    
    \caption{Environments. (\emph{a}) The cascaded double tank system. (\emph{b}) pH-control of wastewater.}%
\end{figure}

\subsection{Cascaded Water Tank Level Control}

The water tank environment consists of two tanks mounted on top of each other, as depicted in Figure~\ref{fig:real-tank-pic}. 
When actuated by the controller, the water from the source is first pumped to the upper tank and then flows to the lower tank through a small hole at the bottom of the upper tank. In the end, the water flows into the container below. 
Using Bernoullis law, a model of the system can be written as
\begin{align}
    \label{eq:env_modelwatertank}
    \begin{split}
    \dot{l}_1(t) &= -\frac{a_1}{A_1}\sqrt{2gl_1(t)} + \frac{K_\text{pump}}{A_1}u(t),\\
    \dot{l}_2(t) &= \frac{a_1}{A_1}\sqrt{2gl_1(t)} - \frac{a_2}{A_2}\sqrt{2gl_2(t)},
    \end{split}
\end{align}where $l_1(t), l_2(t)$ are the water levels of the two tanks, $a_1$, $a_2$ are the areas of the bottom holes, $A_1$, $A_2$ are the areas of the cross sections of the tanks, $K_\text{pump}$ is the pump constant and $u(t)$ is the voltage applied to the pump. 

However, the real system may not follow these equations perfectly, and the exact value of each parameter is not known. Hence, the values of the ratios $p_1 = a_1/A_1$ and $p_2 = a_2/A_2$ are assumed to be in the uniform distribution $\mathcal{U}_1(0.0015, 0.0024)$.
Furthermore, the value of the ratio  $K_\text{pump}/A_1 $ is assumed to be in the uniform distribution $\mathcal{U}_2(0.07, 0.17)$. Thus, the model ensembles of this task can be defined as, $\mathcal{F}_\mathcal{P} = \{ f_p \mid p_1 \in \mathcal{U}_1, p_2 \in \mathcal{U}_1, p_3 \in \mathcal{U}_2 \}$. The task is to learn a robust policy for the cascaded tank system that controls the level of the lower tank $y(t) = l_2(t)$. In the simulation, the system is discretized using Euler's method \cite{glad2018controlbook} with a sampling period of 2 seconds and the reward function at time step $t$ is defined as
\begin{align}
    R(x_t, \yref) = -(y_t - \yref)^2 
    \label{eq:reward_func}
\end{align}
where $x_t$ contains the levels of two tanks and $y_t$ is the level of the bottom tank at time step $t$.

\begin{figure}[ht!]
    \centering
    \subfloat[]{
    \label{fig:res:changing:all}
    \includegraphics[width=0.44\textwidth]{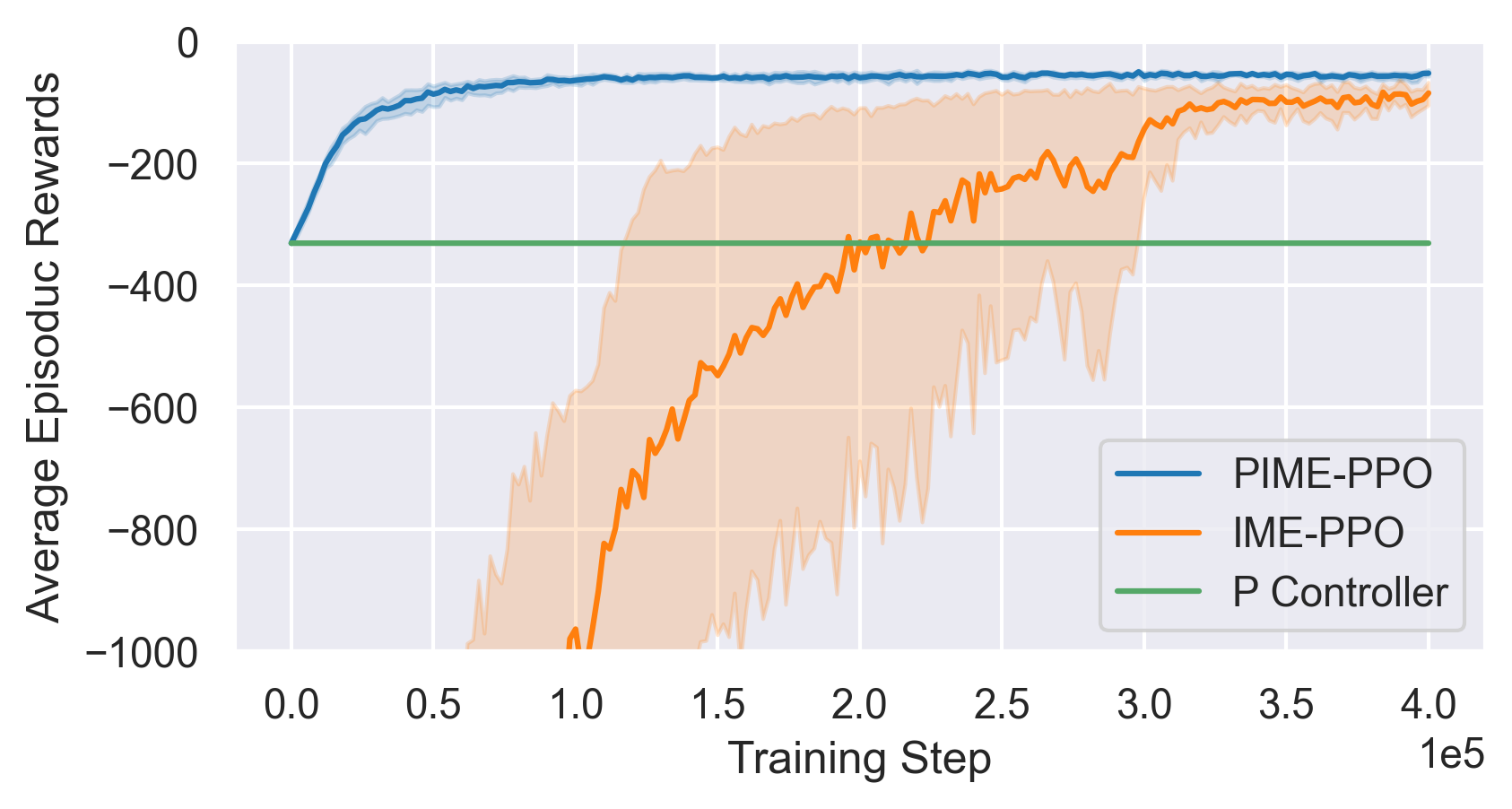}
    }\\
    \centering
    \subfloat[]{
    \label{fig:res:simulation:500}
    \includegraphics[width=0.43\textwidth]{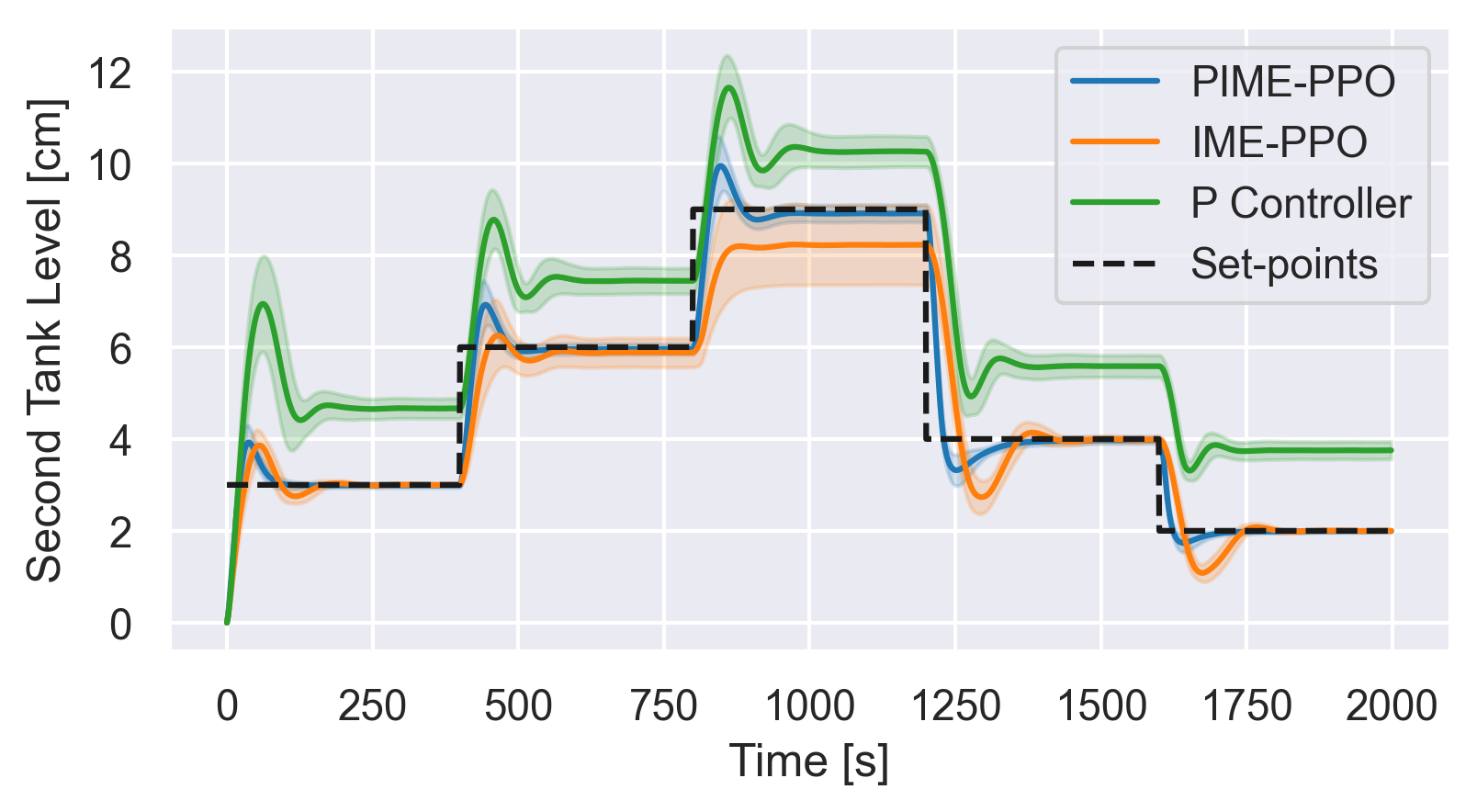}
    }\\
    \subfloat[]{
    \label{fig:res:simulation:seeds}
    \includegraphics[width=0.43\textwidth]{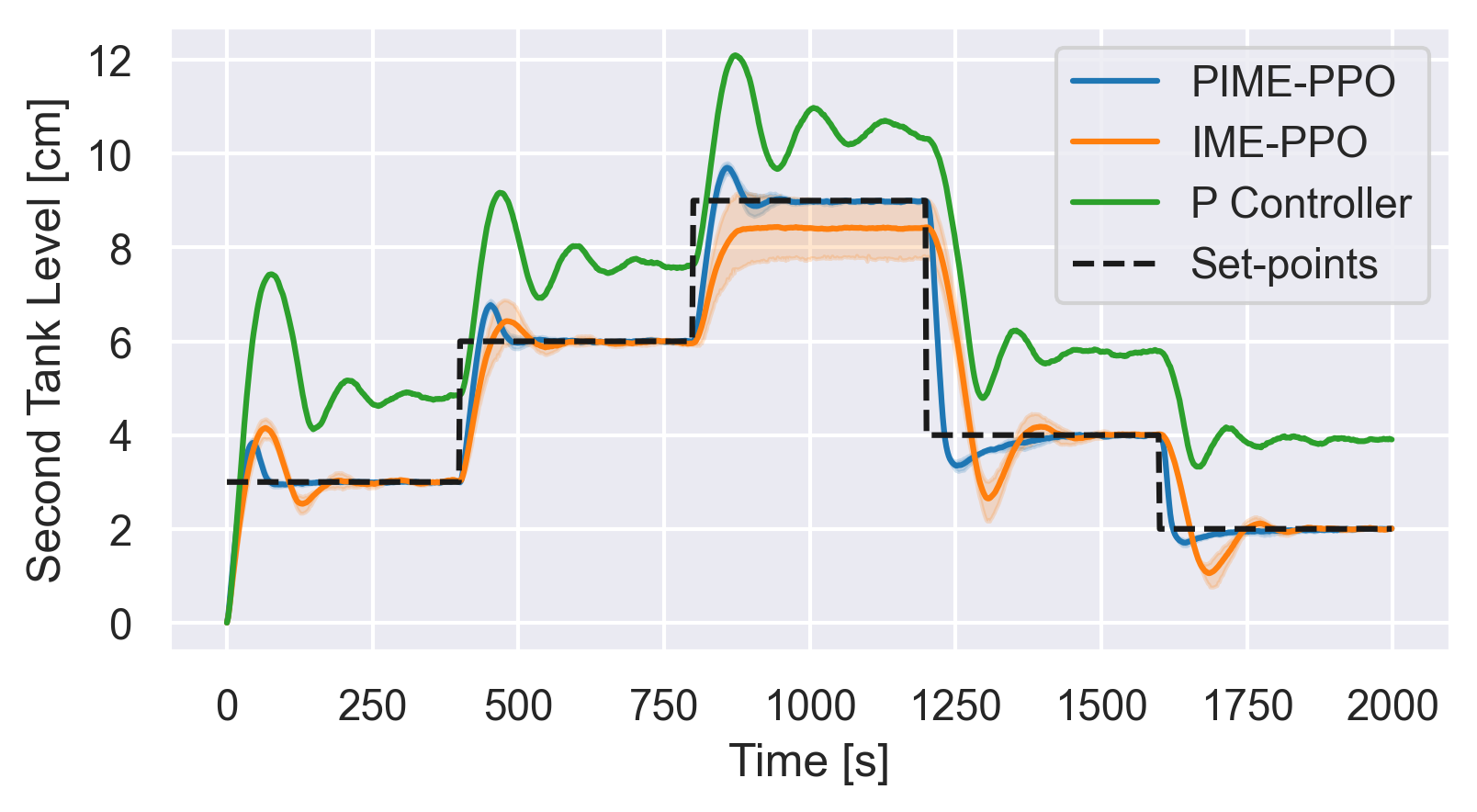}
    }
    
    \caption{Comparison between the prior P-controller, the IME-PPO, and the PIME-PPO in simulation. (\emph{a}) Training performance comparison between the prior P-controller, the IME-PPO and the PIME-PPO for four hundred thousand steps. (\emph{b}) Evaluated trained policies with training random seed $3$ on $500$ models with sampled parameters from $\mathcal{F}_\mathcal{P}$ to track 5 different set-points. (\emph{c}) Evaluated trained policies with five different random seeds on one specific model with sampled parameters to track 5 different set-points.} %
    \label{fig:res:simulations}
\end{figure}

\subsubsection{Question 1: Simulation with Different Model Parameters}
For this example, the prior controller used is an un-tuned P-controller.
In Figure~\ref{fig:res:changing:all}, the training performance of IME-PPO and PIME-PPO are compared. 
The performance of the prior P-controller is also shown to illustrate that the method without the prior controller needs many more samples to train a good policy.   
As shown, the PIME-PPO starts from the P-controller and improves during the entire training without any drop. It converges quickly to its best performance, and the variance between different random seeds is very small. The IME-PPO, on the other hand, starts from scratch and requires more than $2 \times 10^5$ steps before it reaches the performance of the un-tuned P-controller, and it does not manage to reach the performance of PIME-PPO within $4 \times 10^5$ steps. This shows that even an un-tuned prior P-controller can improve the efficiency of the training substantially.

In Figure~\ref{fig:res:simulation:500}, the trained policies are evaluated on 500 different models with different parameters.  
It can be observed that the performance of both IME-PPO and PIME-PPO are better than the un-tuned P-controller. It is also seen that the error bar of PIME-PPO is much smaller than that of IME-PPO which indicates that the policy found by PIME-PPO is more robust to model parameter changes than the one found by IME-PPO. Also, it is found that PIME-PPO overshoots more than IME-PPO. This situation is because the PIME-PPO controller is trained to reach all the levels which of course include the higher levels and in order to minimize the overall tracking error, the policy of PIME-PPO has higher gains before reaching the set-points. 
Finally, it can be noted that PIME-PPO achieves very good tracking performance for all sets of parameters, indicating that it makes good use of the integrated error. 
It should be noted that the error bar of PIME-PPO does not mean that there is a tracking error, but for some sets of parameters, there are oscillations before reaching the set-point. 

In Figure~\ref{fig:res:simulation:seeds}, the five different policies that were trained with different seeds are evaluated on a fixed set of model parameters. Again, it can be observed that the results from PIME-PPO are stable and manage to track the set-points.
However, the IME-PPO has more oscillations and does not manage to track the largest set-points well.

\subsubsection{Question 2: Transfer to the Real-world System}

\begin{figure}[ht]
    \centering
    \subfloat[]{
    \label{fig:res:real:off}
    \includegraphics[width=0.45\textwidth]{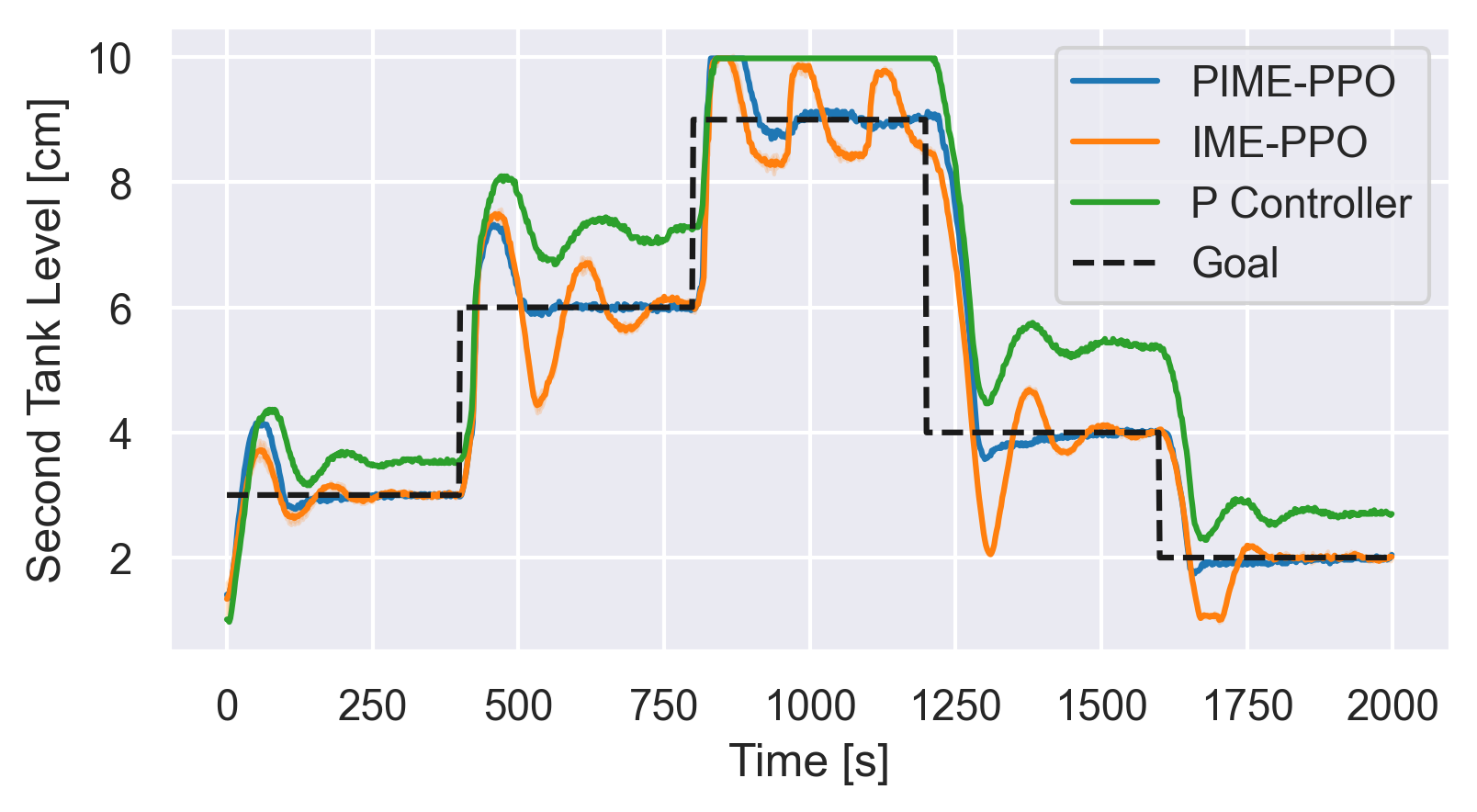}
    } \\
    \subfloat[]{
    \label{fig:res:real:on}
    \includegraphics[width=0.45\textwidth]{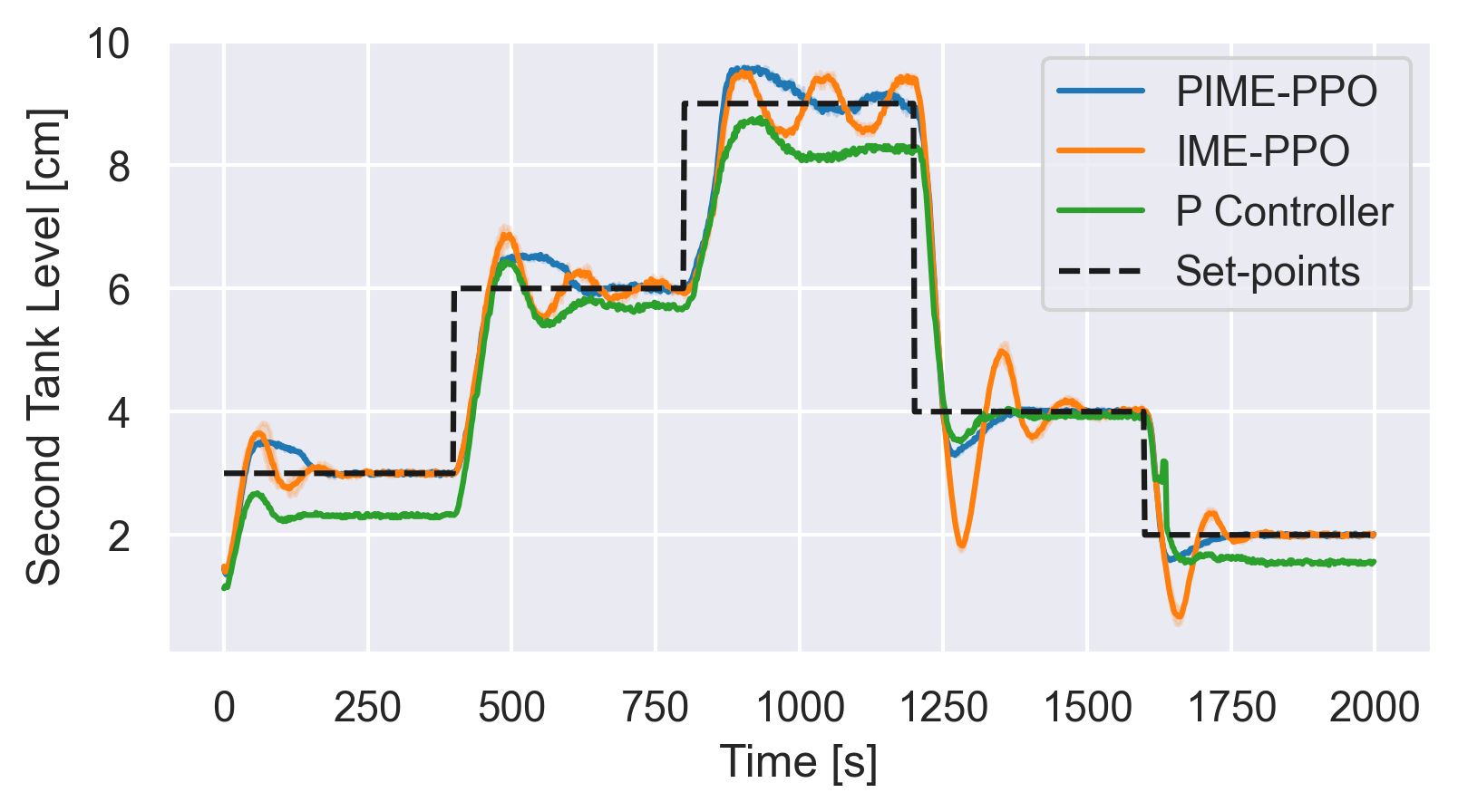}
    }
    
    \caption{Comparisons between the prior P-controller, the IME-PPO, and the PIME-PPO on the real-world cascaded tank system. \emph{(a)} Switch off. \emph{(b)} Switch on.}%
    \label{fig:res:real}
\end{figure}

To recap, the main goal of the paper is to learn a policy in a simulated environment that works directly when deployed to a real-world system. After training, all policies are evaluated on the real cascaded tank system. Consistent with the simulation, the sampling period is also $2$ seconds. This system differs from the simulation since the tanks may overflow and the sensors are noisy. 
The result of the evaluation of the policy, when applied to the real-world system, is shown in Figure~\ref{fig:res:real:off}. As shown, the residual policies still manage to track different set-points correctly. Compared to the IME-PPO, the PIME-PPO performs better for all set-points with smaller overshoots and oscillations.

\subsubsection{Question 3: Unmodelled effects}
Finally, an unmodelled disturbance is introduced by opening a switch so that water can flow out of the upper tank directly into the container. The result is shown in  Figure~\ref{fig:res:real:on}. It can be seen that, even with such unmodelled effects, both RL methods can reach and track the set-points thanks to the integrator.

\subsection{pH-Control}
pH-control is a challenging control problem in the chemical industry since the process includes chemical equilibrium, kinetics, and mixing dynamics that together result in a highly nonlinear dynamic system that is notoriously hard to control well. One important issue is that the process buffer capacity varies over time \cite{syafiie2007phmodel}. Typically the buffer capacity is unknown and changes the gain of the system significantly, making it even harder to control. 
The pH-control of wastewater containing ammonia (NH3) and sodium hydroxide (NaOH) is  considered here, stated as a continuous neutralization titration process by hydrochloric acid (HCl). The process is illustrated in Figure~\ref{fig:ph-pic}. 
The chemical reaction equilibrium of the environment can be stated using a cubic equation for $[H^+]$ \cite{wigren1990wiener},
\begin{align}
     [H^+]^3 &+ ([NH_3] - [HCl]+ [NaOH] + K)[H^+]^2  \\ 
     &+ (K[NaOH] - K[HCl] - KK_w)=0,
\end{align}
where $[NH_3], [HCl],[NaOH]$ are the concentration of the substances, and where $K$ and $K_w$ are the constants of the reaction. 
Using mass balances, it's straightforward to obtain the following ODE that describes the mixing dynamics
\begin{equation}
\dot{[HCl]}(t)=-\frac{q_{ww}}{V}[HCl](t)+\frac{[HCl,C]q_{[HCl,C]}}{V}u(t).
\label{new}
\end{equation}
Here $q_{ww}$ is the inlet flow, $V$ is the reactor tank volume, $[HCl,C]$ is a constant representing the concentration of the hydrochloric acid used for control, $q_{[HCl,C]}$ represents a constant control flow, and $u(t)$ is hence a relative quantity and not the flow itself. The mixing results in concentrations in the tank as before reactions take place, where the state is the hydrochloric acid concentration after mixing.  
The pH value at time step $t$ can be calculated with
$x(t) = -log_{10}[H^+](t)$.
 In the simulation, the system is discretized using Euler's method \cite{glad2018controlbook} with a sampling period of 20 seconds and the reward function at time step $t$ is similar to \eqref{eq:reward_func} and defined as
\begin{align}
    R(x_t, \yref) = -(x_t - \yref)^2.
\end{align}
where $x_t$ is the pH level of time step $t$.
Also, in order to deal with the capacity uncertainty in terms of the inlet and outlet flows, the model ensemble is defined as 
\begin{align}
    \mathcal{F}_{\mathcal{P}} = \{&p_1=\frac{[HCl,C]q_{[HCl,C]}}{V}\sim \mathcal{U}_3(0.005,0.015), \\
    &p_2=\frac{q_{ww}}{V}\sim \mathcal{U}_4(0.0015, 0.0025)\}.
\end{align}

\begin{figure}[ht]
    \centering
    \includegraphics[width=0.4\textwidth]{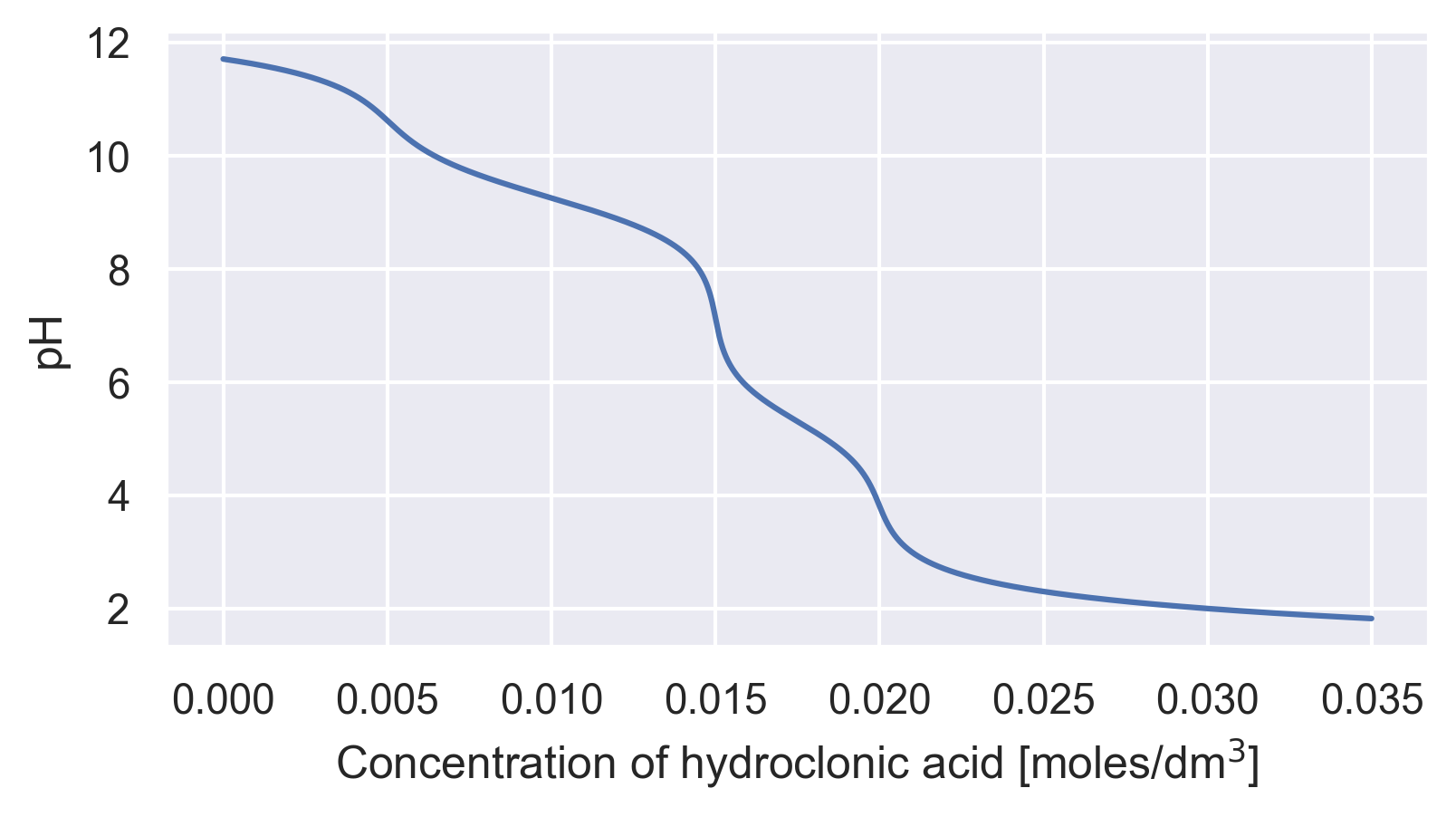}
    \caption{The relationship between the concentration of hydrochloric acid ($[HCl]$) and the pH value with $[NH_3]=0.01~\text{moles/dm}^3,[NaOH]=0.01~\text{moles/dm}^3$ and concentration of weak acid $[H^+]=0.005~\text{moles/dm}^3$.}
    \label{fig:ph-nonlinear}
\end{figure}

Due to the strong nonlinearity of the system shown in Figure~\ref{fig:ph-nonlinear}, the control task is challenging.

\subsubsection{Question 1: Simulation with Different Model Parameters}
\begin{figure}[h!]
    \centering
    \subfloat[]{
    \label{fig:res:ph:training}
    \includegraphics[width=0.44\textwidth]{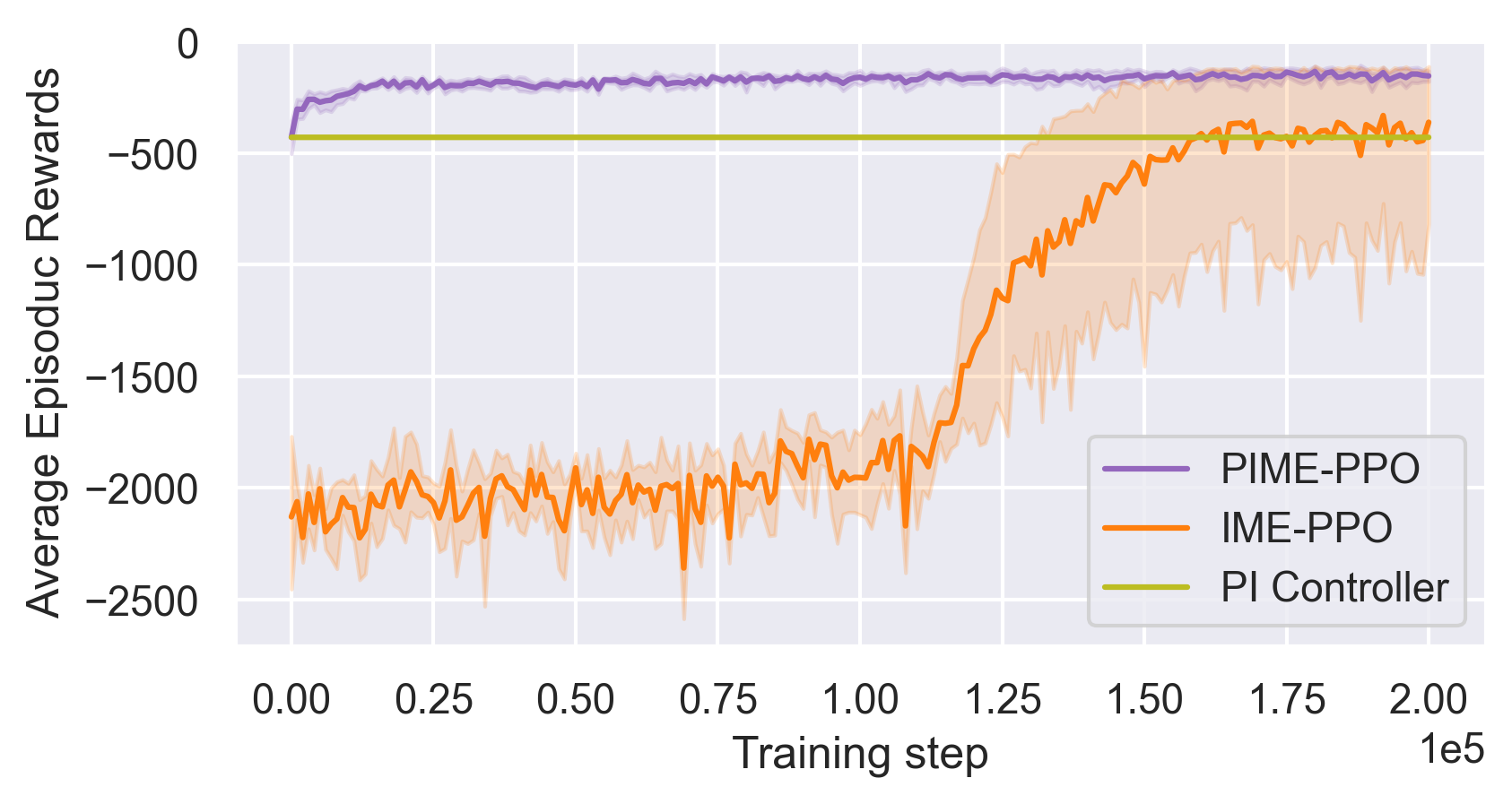}
    }\\
    \centering
    \subfloat[]{
    \label{fig:res:ph:500}
    \includegraphics[width=0.43\textwidth]{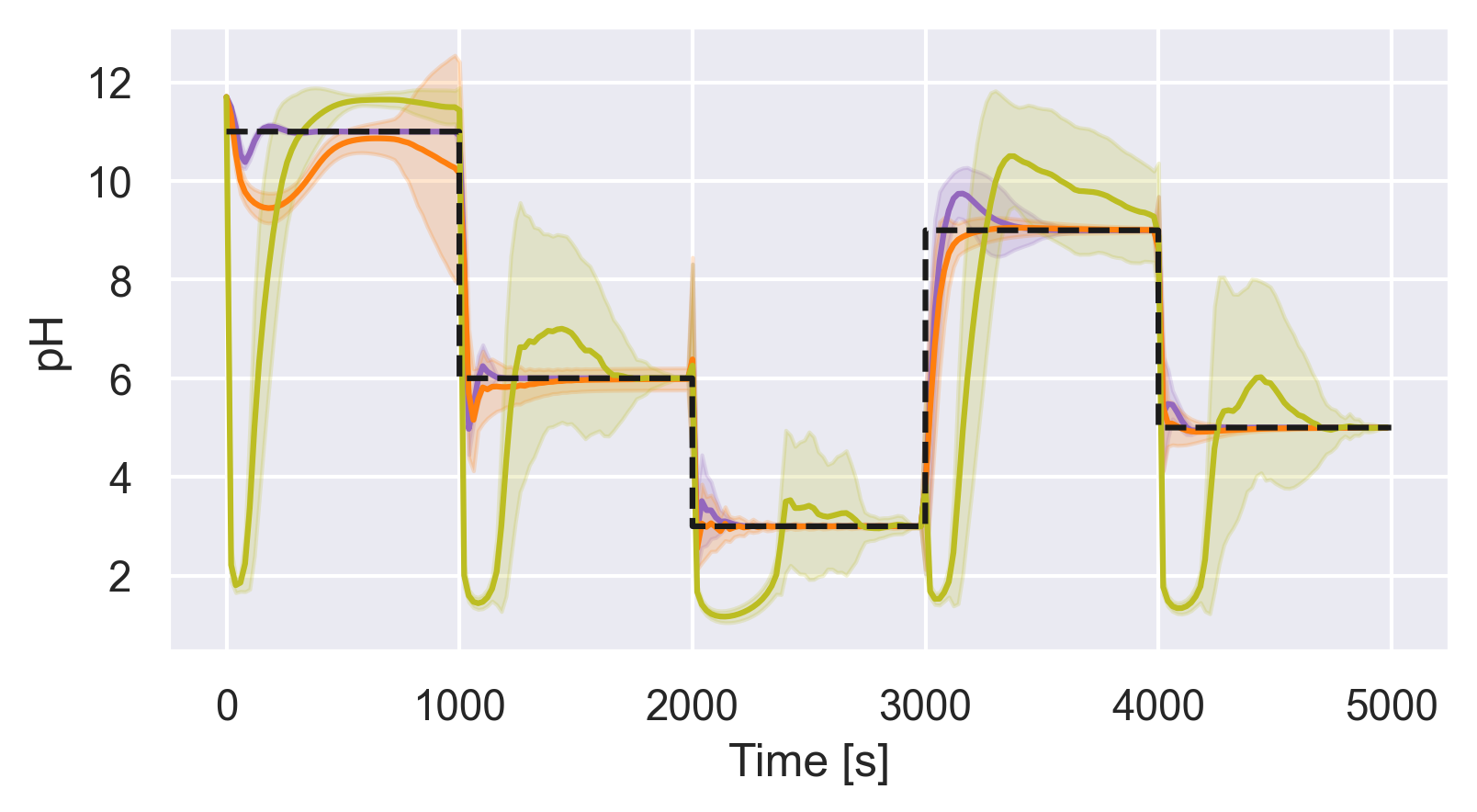}
    }\\
    \subfloat[]{
    \label{fig:res:ph:seeds}
    \includegraphics[width=0.43\textwidth]{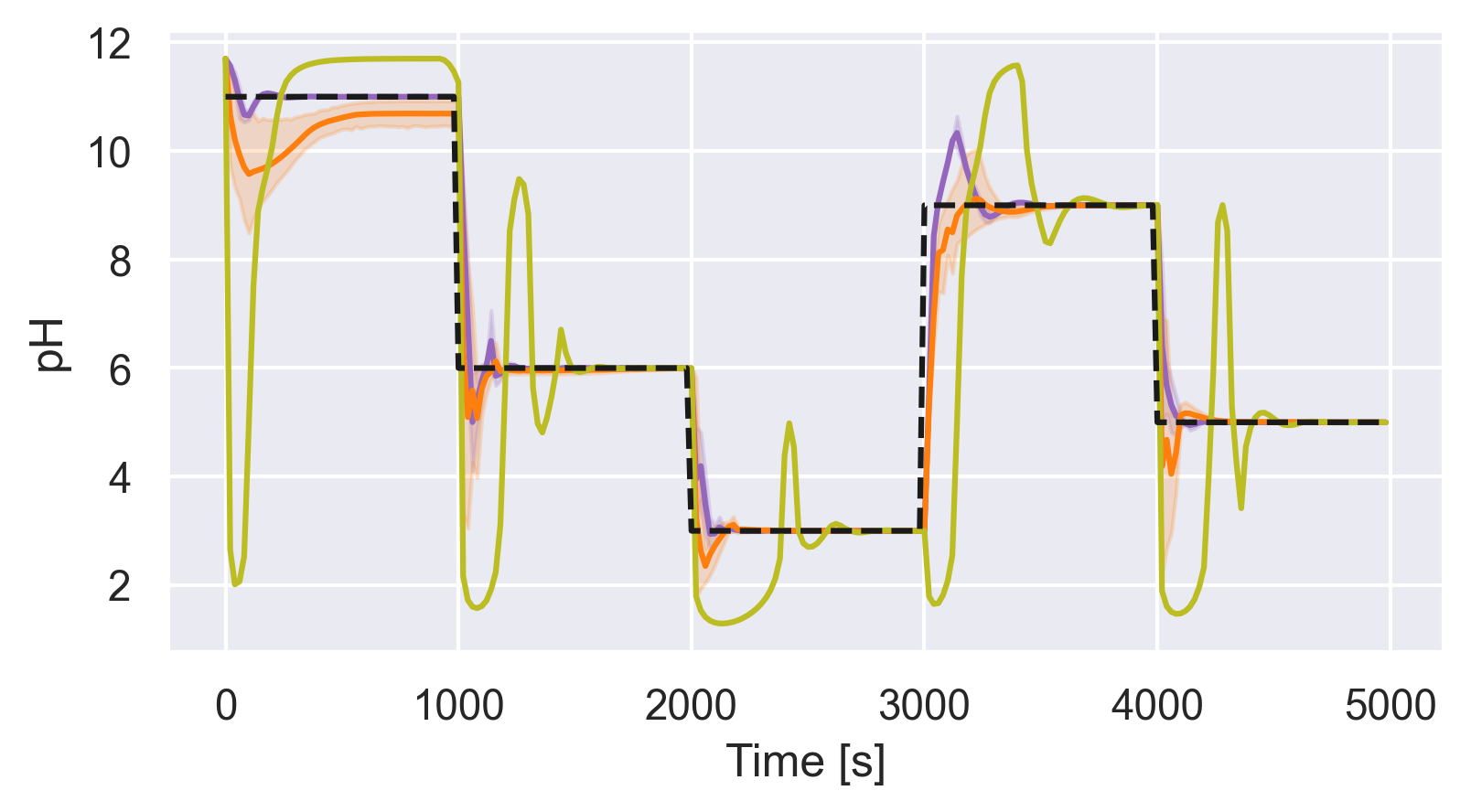}
    }
    
    \caption{Comparison between the prior P-controller, the IME-PPO, and the PIME-PPO in simulation to track 5 different set-points. (\emph{a}) Training performance comparison between the prior PI-controller, the IME-PPO and the PIME-PPO in simulation for four hundred thousand steps. (\emph{b}) Evaluated trained policies with training random seed $3$ on $500$ models with sampled parameters from $\mathcal{F}_\mathcal{P}$. \emph{Black} line is the set-points level. (\emph{c}) Evaluated trained policies with five different random seeds on one specific model with sampled parameters.}%
    \label{fig:res:ph}
\end{figure}

An un-tuned PI-controller is used as the prior controller. 
In Figure~\ref{fig:res:ph:training}, the training performance of IME-PPO and PIME-PPO are depicted together with the prior PI-controller.
As shown, the PIME-PPO starts from the PI-controller and improves during the entire training session, exactly as for the cascaded tank example. The IME-PPO, on the other hand, remains locked until $10^5$ steps have passed, when it starts to explore better and converge. The final value of IME-PPO appears to be worse than for PIME-PPO.

In Figure~\ref{fig:res:ph:500}, the trained policies are evaluated on 500 different models with different parameters. Again, IME-PPO and PIME-PPO improve substantially as compared to the PI-controller which has large overshoots.

In the experiment, the integrator in PIME-PPO was bounded to avoid the accumulation of too-large integrated errors during training and evaluation.
The policy found by PIME-PPO responds better to set-point changes than IME-PPO. Finally, it can be noted that PIME-PPO achieves very good stationary tracking for all sets of parameters, indicating that it makes good use of the integrated error. Here,  the variance of PIME-PPO originates from overshoots before reaching the set-point.

In Figure~\ref{fig:res:ph:seeds}, the five different policies that were trained with different seeds are evaluated on
a fixed set of model parameters. Again, it can be observed that the results of PIME-PPO are stable and manage to track the set-point. Without the prior PI-controller, the IME-PPO has a slower response for high-level set-points where the gain of the pH control system is low, cf. Figure~\ref{fig:ph-nonlinear}.

\section{Conclusion}
\label{sec:conclusion}
The paper discussed the integration of machine learning and control-theoretic methods, to obtain a new multi-goal RL method capable of robust set-point control of real-world nonlinear systems. More precisely, simulated policy optimization over a model ensemble was used to enhance robustness with respect to varying dynamics in general. To achieve set-point tracking, integrating control was included by means of an integrated error state trained with a neural network operating in parallel with the network for the error state itself. Here the use of model ensemble optimization is crucial to avoid the training replacing the integrated error with an exact gain, that would be the case if a single model would be trained. Finally, exploration in the amplitude domain was added, using a simple aiding feedback controller, like a proportional controller. This was shown to reduce the training time significantly. The new method was evaluated on a  cascaded tank system, proving both  robustness and exact set-point tracking. The method was also tested on a strongly nonlinear pH-control system, also with good results.  Interesting remaining tasks include a theoretical study of the stability properties of the closed-loop system.

\bibliographystyle{IEEEtran} %

\bibliography{pime.bib}

\newpage
\appendix
\section{Appendix}

\subsection{Hyperparameters}
\label{appendix:hyper}
The hyperparameters used  in the water tank level control task and pH control task are shown in Table~\ref{table:hyper}. In each task, all the experiments share the same hyperparameters.

\begin{table}[ht]
    \centering
    \begin{tabular}{lll}
     Hyperparameter & Water Tank & pH \\ \hline
     Horizon & 200  & 50  \\
     Adam stepsize & $3\times10^{-4}$  & $3\times10^{-4}$ \\
     Minibatch size & 256& 128 \\
     Number of epochs   & 10 & 40 \\
     Discount $\gamma$ & 0.995 & 0.98\\
     GAE parameter $\lambda$ & 0.97  & 0.97 \\
     Clipping parameter $\epsilon$ &0.2 & 0.2 \\
     PPO Value Function coefficient $c_1$ &  1.0 &  1.0 \\
     PPO Value Function coefficient $c_2 $ &  0.02 &  0.02 \\
      Integrator bounds & [-25,25]  & [-25,25] \\
      Number of episodes collected per iteration $M$ & 5 & 5 \\\\
    \end{tabular}
    \caption{Hyperparameters used in the water tank level control task.}
    \label{table:hyper}
\end{table}

\subsection{Modular Neural Network}
\label{appendix:nn}
\begin{figure}[ht]
    \centering
    \begin{tikzpicture}[thick,scale=0.7, every node/.style={scale=0.8},shorten >=0.6pt,->,draw=black!40, node distance=0.7*\layersep]
    \tikzstyle{every pin edge}=[<-,shorten <=0.6pt]
    \tikzstyle{neuron}=[circle,fill=black!25,minimum size=10pt,inner sep=0pt]
    \tikzstyle{input neuron}=[neuron, fill=green!40];
    \tikzstyle{output neuron}=[neuron, fill=red!40];
    \tikzstyle{hidden neuron}=[neuron, fill=blue!40];
    \tikzstyle{annot} = [text width=2em, text centered]

    Draw the input layer nodes
    \node[input neuron, pin=left: $x_t^{(1)}$] (I-1) at (0,-1) {};
    
    \node[input neuron, pin=left: $\vdots~~$] (I-2) at (0,-2) {};
    \node[input neuron, pin=left: $x_t^{(n)}$] (I-3) at (0,-3) {};

    \node[input neuron, pin=left: $\yref$] (I-4) at (0,-4) {};
    \node[input neuron, pin=left: $z_t$] (I-5) at (0,-5) {};

    \foreach \name / \y in {1,...,5}
        \path[yshift=0.3cm]
            node[hidden neuron] (H-\name) at (\layersep,-\y cm) {};
    
    \foreach \name / \y in {4,...,6}
        \path[yshift=0.3cm]
            node[hidden neuron] (H-\name) at (\layersep,-\y cm) {};
    
    \foreach \name / \y in {1,...,6}
        \path[yshift=0.3cm]
            node[hidden neuron] (R-\name) at (\layersep*2,-\y cm) {};
    
     \foreach \name / \y in {1,...,6}
        \path[yshift=0.3cm]
            node[hidden neuron] (G-\name) at (\layersep*3,-\y cm) {};

    \node[output neuron,pin={[pin edge={->}]right:$u_t$}, right of=G-3] (O) {};

    \foreach \source in {1,...,4}
        \foreach \dest in {1,...,4}
            \path (I-\source) edge (H-\dest);
    
    \foreach \source in {5}
        \foreach \dest in {5,...,6}
            \path (I-\source) edge (H-\dest);
            
    \foreach \source in {1,...,4}
        \foreach \dest in {1,...,4}
            \path (H-\source) edge (R-\dest);
        
     \foreach \source in {5,...,6}
        \foreach \dest in {5,...,6}
            \path (H-\source) edge (R-\dest);
            
    \foreach \source in {1,...,6}
        \foreach \dest in {1,...,6}
            \path (R-\source) edge (G-\dest);
            
    \foreach \source in {1,...,6}
        \path (G-\source) edge (O);
    \node[annot,above of=H-1, node distance=1cm] (hl) {};
    \node[annot,left of=hl] {};
    \node[annot,right of=hl] (hl2) {};
    \node[annot,right of=hl2] (hl3) {};
    \node[annot,right of=hl3] (hl3) {};
\end{tikzpicture}
    \caption{Network architecture of $g_\theta$ in PIME-PPO. Bias and activation functions are ignored.}
    \label{fig:network}
\end{figure}
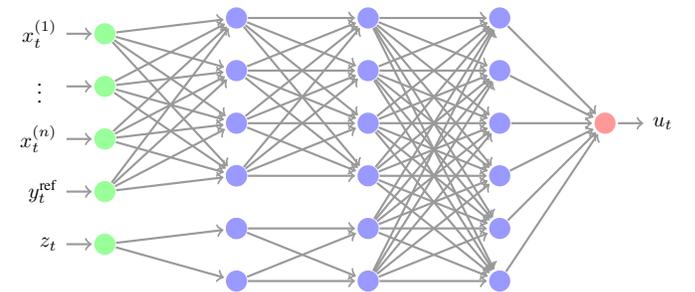

As described in Section~\ref{sec:exp}, the $g_\theta$ of PIME-PPO is a modular neural network with separate networks for the integrated error $z_t$ and the other items $x_t$ and $\yref$, see Figure~\ref{fig:network}. 

\end{document}